\documentstyle[preprint,aps]{revtex}
\tightenlines
\begin{document}
\title{Nucleon-Nucleon Interactions and Observables}
\author{Wayne N. Polyzou}
\address{ 
Department of Physics and Astronomy, 
The University of Iowa, 
Iowa City, IA 52242
}
\date{\today}
\maketitle
\footnotetext{P.A.C.S. 13.175c, 21.60} 
\begin{abstract}

A class of nucleon-nucleon interactions which are exactly phase
equivalent to a given realistic nucleon-nucleon interaction are
exhibited. These interaction have the property that the RMS radius of
the deuteron can be made arbitrarily large without changing any of the
scattering or bound-state observables.  With this construction it is
possible to find realistic interactions that do not obey the linear
relation between the RMS radius and the triplet scattering length
observed by Klarsfeld et. al. \cite{Klarsfeld}.   The interpretation of
these examples is discussed.

\end{abstract}
\newpage
\narrowtext
\section{Introduction}

In 1986 Klarsfeld, Martorell, Oteo, Nishimura, and Sprung
\cite{Klarsfeld} observed an empirical linear relation between the
triplet scattering length and the deuteron RMS radius in a large class
of realistic nucleon-nucleon interactions.  In this paper I exhibit
sequences of realistic nucleon-nucleon interactions that have identical
S-matrix elements, identical bound-state observables, and arbitrarily
large deuteron RMS radius.  In addition, for each interaction there is an
exchange current that leaves all electromagnetic observables unchanged.

The construction of the model interactions is based on two observations.
The first is that unitary transformations that are compact perturbations
of the identity leave all scattering and bound state observables
unchanged.  The second that $r^2$ is an unbounded operator.  Using these
two observations it is possible to construct model interactions with the
desired properties.  The resulting model interactions differ from the
given initial nucleon-nucleon interaction on a two-dimensional
subspace of the Hilbert space.

The key result of this paper is the following theorem:

\bigskip

\noindent{\bf Theorem:} Let $V:=V_0$ be a realistic nucleon-nucleon 
interaction with Hamiltonian $H=H_0=K+V_0$.   Let $r_0$ be the deuteron 
RMS radius in this model.  There exist an infinite sequence of
interactions $\lbrace V_m\rbrace_{m=1}^{\infty}$ and two-body
Hamiltonians $H_m=K+ V_m$ with the properties

\begin{enumerate}
\item Each $H_m$ has the same spectrum and scattering observables as $H_0$.

\item If $\vert \psi_m \rangle$ is the deuteron eigenstate of $H_m$ then 
\begin{equation}
<r^2>_m := \langle \psi_m \vert r^2 \vert \psi_m \rangle \geq mr_0^2.
\label{eq:AA}
\end{equation}
\end{enumerate}

\section{Proof of the Theorem}

In all that follows I assume that all operators are considered to 
have domain and
range in the relative motion Hilbert space ${\cal H}_{rel}$ defined by:
\begin{equation}
{\cal H } = {\cal L}^2 (R^3, d^3p) \otimes {\cal H}_{rel}.
\label{eq:AAB}
\end{equation} 

To prove the theorem I first exhibit a class of unitary transformations that
preserve all bound state and scattering observables.   
To do this let $\vert \chi \rangle$ be any unit normalized vector and define
\begin{equation}
W := I + {2 i \over 1-i} \vert \chi \rangle \langle \chi \vert .
\label{eq:BA}
\end{equation}
Direct computation shows that 
\begin{equation}
W^{\dagger}W = I.
\label{eq:BB}
\end{equation}
 
Next note that if \cite{Coester}
\begin{equation}
\lim_{t\to \pm \infty} \Vert (W-I ) e^{-iKt} \vert \phi \rangle \Vert =0
\label{eq:BC}
\end{equation}
for every vector $\vert \phi \rangle \in {\cal H}_{rel}$ and {\it both} time
limits then it follows that 
\[
0= \lim_{t \to \pm \infty} \Vert e^{iHt}(W-I) e^{-iKt} \vert \phi \rangle \Vert   =
\]
\[
\lim_{t \to \pm \infty} \Vert (W e^{iW^{\dagger} HWt} - e^{iHt} ) 
e^{-iKt} \vert \phi \rangle \Vert   =
\]       
\begin{equation}
\Vert [W \Omega_{\pm} (W^{\dagger} H W) - \Omega_{\pm} (H)]  
\vert \phi \rangle \Vert .  
\label{eq:BD}
\end{equation}
With $H'$ defined by 
\begin{equation}
H':= W^{\dagger}HW 
\label{eq:BE}
\end{equation}
equation (\ref{eq:BC}) implies the M\o ller wave operators for $H$ and
$H'$ are related by  
\begin{equation}
\Omega_{\pm} (H) = W \Omega_{\pm}(H') 
\label{eq:BF}
\end{equation}
which is valid for both asymptotic conditions.  It follows that the 
scattering operators $S$ and $S'$ for $H$ and $H'$ are identical:
\[
S:= \Omega_+^{\dagger}(H) \Omega_-(H) = 
\]
\begin{equation}
\Omega_+^{\dagger}(H')W^{\dagger} W \Omega_-(H') =
\Omega_+^{\dagger} (H') \Omega_- (H') = S'.
\label{eq:BG}
\end{equation}
Note that the free relative kinetic energy $K$ is {\it not} transformed 
in the wave operator.  

This shows that any unitary transformation $W$ satisfying (\ref{eq:BC}) will
preserve the spectrum and all scattering observables.  The resulting
transformed Hamiltonian has the form

\begin{equation}
H' = H + {2i \over 1-i} \vert \chi \rangle \langle
\chi \vert H 
- {2i \over 1+i} H \vert \chi \rangle \langle \chi \vert +
2 \vert \chi \rangle \langle 
\chi \vert H \vert \chi \rangle \langle \chi \vert 
\label{eq:BH}
\end{equation}
which differs from $H$ only on the two dimensional subspace spanned by
$\vert \chi \rangle$ and $H\vert \chi \rangle$.
 
To establish the desired equivalence it is sufficient to prove 
(\ref{eq:BC}) for $W$ of the form (\ref{eq:BA}).  This 
is equivalent to establishing
\begin{equation}
\lim_{t \to \pm \infty} \vert \langle \chi \vert e^{-iKt} 
\vert \psi \rangle \vert =0 .
\label{eq:BI}
\end{equation}
Inserting intermediate eigenstates of $K$, realizing that $K$ has
absolutely continuous spectrum, gives:
\[
\lim_{t \to \pm \infty} \vert \int_0^{\infty}  
e^{-i k \cdot t } \int dx \langle \chi \vert k,x \rangle 
\langle k,x \vert \psi \rangle \vert =
\]
\begin{equation}
\lim_{t \to \pm \infty} \vert  \int_0^{\infty} dk f(k)   
e^{-i k \cdot t } \vert  =0.
\label{eq:BJ}
\end{equation}
Since 
\begin{equation}
f(k) := 
\int dx \langle \chi \vert k,x \rangle \langle k,x \vert \psi \rangle 
\label{eq:BK}
\end{equation}
is a product of square integrable functions of $k$ it is an  $L^1$
function of $k$ after integrating out the other observables $x$.   The
vanishing of (\ref{eq:BJ}) is a consequence of the Riemann-Lebesgue
lemma \cite{Katznelson}.

This establishes that all Hamiltonians $H'$ of the form (\ref{eq:BH}) have
the same bound state and scattering observables as the initial
Hamiltonian $H$.

To complete the proof of the theorem I show that it possible, by a
careful choice of  the vectors $\vert \chi \rangle$, to make the 
deuteron RMS radius as large as desired.  To begin the construction first
note that the deuteron RMS radius for the $H'$ Hamiltonian is 
\[
\langle r^2 \rangle ' := \langle \psi' \vert r^2 \vert \psi' \rangle =
\langle \psi \vert W r^2 W^{\dagger} \vert \psi \rangle =
\]
\begin{equation}
\langle \psi \vert (1 + {2i \over 1-i} \vert \chi \rangle \langle \chi
\vert ) r^2  (1 - {2i \over 1+i} \vert \chi \rangle \langle \chi \vert
)\vert \psi \rangle .
\label{eq:BL}
\end{equation}
Choose $\vert \chi \rangle $ so it satisfies 
\begin{equation}
\langle \chi \vert r^2 \vert \psi \rangle = 
\langle \psi \vert r^2 \vert \chi \rangle =0
\label{eq:BM}
\end{equation}
and
\begin{equation}
\langle \chi \vert \psi \rangle  \not= 0 .
\label{eq:BN}
\end{equation}
With this choice the cross terms in (\ref{eq:BL}) do not contribute which 
gives the following expression for the deuteron RMS radius: 
\begin{equation}
\langle r^2 \rangle '= \langle \psi \vert r^2 \vert \psi \rangle +
2 \vert \langle \psi \vert \chi \rangle \vert^2  
\langle \chi \vert r^2 \vert \chi \rangle .
\label{eq:BO}
\end{equation}
The theorem is proved by exhibiting a sequence of $\vert \chi_m \rangle$
satisfying (\ref{eq:BM}-\ref{eq:BN}) with the  property that  $\vert
\langle \psi \vert \chi_m \rangle \vert$ is bounded from below, and
$\langle \chi_m \vert r^2 \vert \chi_m \rangle$ is arbitrarily large.

To construct such a sequence of vectors let 
\begin{equation}
\vert \xi_0 \rangle = N r^2 \vert \psi \rangle  
\label{eq:BP}
\end{equation}
where $N$ is a constant chosen to make this vector normalized 
to unity.  Let $\lbrace 
\vert \xi_m \rangle \rbrace_{n=0}^{\infty} $ any complete orthonormal basis 
in the domain of $r^2$ with $\vert \xi_0\rangle$ as 
the first vector and $\vert \xi_1 \rangle$
satisfying 
\begin{equation}
\langle \xi_1  \vert \psi \rangle = c =c^* > 0 .
\label{eq:BQ}
\end{equation}

Note the following properties:
\begin{equation}
\lim_{n\to \infty} \langle \xi_n \vert \psi \rangle =0
\label{eq:BR}
\end{equation}
\begin{equation}
\lim_{n\to \infty} \langle \xi_n \vert r^2 \vert \xi_1 \rangle =0 
\label{eq:BS}
\end{equation}
and for each finite $R>0$ there is an $n$ such that 
\label{eq:BT}
\begin{equation}
\langle \xi_n \vert r^2 \vert  \xi_n \rangle > R .
\label{eq:BU}
\end{equation}
The first two properties are required for $\vert \phi \rangle$ and 
$r^2 \vert \phi \rangle$  be normalizable.  If the last property
is a mathematical form of the statement that $r^2$ is unbounded. 

What these properties mean is given any arbitrarily small $\epsilon >0$ and
any finite $R>0$ it is
possible to find an $n=n(\epsilon , R)$ such that 
\begin{equation}
\vert \langle \xi_n \vert \psi \rangle \vert < \epsilon
\label{eq:BV}
\end{equation}
\begin{equation}
\vert \langle \xi_n \vert r^2 \vert \xi_1 \rangle \vert < \epsilon r_0^2 
\label{eq:BW}
\end{equation}
\begin{equation}
\vert \langle \xi_n \vert r^2 \vert  \xi_n \rangle \vert > R .
\label{eq:BX}
\end{equation}
Let $m$ be a given integer, choose $n$ so 
$\epsilon = .01 c$, and  $R$ satisfying
\begin{equation}            
R > {2(m-1) r^2_0 \over  (.99 c)^2 } + .02 c r_0^2 - r_1^2 .
\label{eq:BY}
\end{equation}
where
\begin{equation}
r_1^2 := \langle \xi_1 \vert r^2 \vert \xi_1 \rangle .
\label{eq:BAD}
\end{equation}

For this choice of $m$ define 
\begin{equation}
\vert \chi_m \rangle := 
{1 \over \sqrt{2}}(\vert \xi_1 \rangle + \vert \xi_n \rangle ).
\label{eq:BZ}
\end{equation}
By construction 
\begin{equation}
\langle \xi_n \vert \psi \rangle = \epsilon_m  e^{i \phi_m}
\qquad \epsilon_m \leq \epsilon  
\label{eq:BAB}
\end{equation}
\begin{equation}
\langle \xi_n \vert r^2 \vert \xi_1 \rangle = r_0^2 \epsilon'_m e^{i \phi_m'}
\qquad \epsilon'_m \leq \epsilon .
\label{eq:BAC}
\end{equation}

It follows that 
\[
\langle \psi_m \vert r^2 \vert \psi_m \rangle =
\]
\[
r_0^2 + 2 \vert \langle \psi \vert \chi_m \rangle \vert^2 \langle \chi_m \vert
r^2 \vert \chi_m \rangle =
\]
\begin{equation}
r_0^2 + {1 \over 2} \vert c^2+ 2 \epsilon_m \cos (\phi_m) + \epsilon_m^2 \vert 
[r^2_1 + 2 r_0^2 \epsilon_m' \cos
(\phi_m') + \langle \xi_n \vert r^2 \vert \xi_n \rangle ].
\label{eq:BAE}
\end{equation}
The relations (\ref{eq:BQ}), (\ref{eq:BV}-\ref{eq:BY}), and 
(\ref{eq:BAB}-\ref{eq:BAC}) imply 
\begin{equation}
\vert c^2+ 2 \epsilon_m \cos (\phi_m) + \epsilon_m^2 \vert 
> \vert c-\epsilon \vert^2 = (.99c)^2
\label{eq:BAF}
\end{equation}
and 
\[
[
r_1^2  + 2 r_0^2 \epsilon_m' \cos
(\phi_m') + \langle \xi_n \vert r^2 \vert \xi_n \rangle ]
\geq
\]
\begin{equation} 
r_1^2 - 2 r_0^2 \epsilon + R = r_1^2 - .02 c r_0^2  + R .
\label{eq:BAG}
\end{equation} 

Putting these together gives 
\[
\langle \psi_m \vert r^2 \vert \psi_m \rangle \geq
\]
\begin{equation}
r_0^2 + {1 \over 2} (.99c)^2 ( r_1^2 - .02 c r_0^2  + R ).
\label{eq:BAH}
\end{equation}
Choosing $R$ satisfying (\ref{eq:BY}) gives
\begin{equation}
\langle \psi_m \vert r^2 \vert \psi_m \rangle \geq  mr_0^2 
\label{eq:BAJ}
\end{equation}
which completes the proof of the theorem.  

The relevant interactions have the general form
\[
V_m =
\]
\begin{equation}
V +  {2i \over 1-i} \vert \chi_m \rangle \langle
\chi_m \vert H 
- {2i \over 1+i} H \vert \chi_m \rangle \langle \chi_m \vert +
2 \vert \chi_m \rangle \langle 
\chi_m \vert H \vert \chi_m \rangle \langle \chi_m \vert .
\label{eq:BAK}
\end{equation}
Since each of these interactions all have the same scattering operator
they all have the same triplet scattering length.  The deuteron RMS
radius can be made as large as desired by choosing $m$ sufficiently
large.  Note that the radius can be adjusted continuously from $r_m$ to
$r_{m+1}$ by interpolating between $\vert \chi_m \rangle$ and $\vert
\chi_{m+1} \rangle$ using 
\begin{equation}
\vert  \chi \rangle  \to {\lambda \vert \chi_m \rangle + (1 -
\lambda )\vert \chi_{m+1} \rangle  \over \sqrt{2 \lambda^2 - 2 \lambda + 1}} .
\label{eq:BAL}
\end{equation}

\section{Interpretation}

The interactions constructed in this paper show by explicit example that
an interaction fit to bound state and scattering data need not have any
relation between triplet scattering length as deuteron RMS radius.
The models constructed all have a separable nonlocality that
acts on a two-dimensional subspace.  

This observation is related to the fact that the RMS radius is not an
observable.  The RMS radius should be distinguished from the charge
radius which is a scattering observable.  In the models constructed in
this paper the hadronic current must also be transformed with the same
unitary transformation.  If this is done the electromagnetic observables
remain unchanged, however this is achieved  by introducing exchange
currents.  Specifically given  the hadronic current operator in the $0$
representation,  the current in the $m$-representation is related by
\begin{equation}
J_m^{\mu} (x) = W_m^{\dagger} J_0^{\mu}(x ) W_m   
\end{equation}
or
\begin{equation}
J_0^{\mu} (x) = W_m J^{\mu}(x ) W^{\dagger}_m.   
\end{equation}
The two-body parts of the unitary operators $W_m$ generate 
exchange current contributions.

The result of the construction in this paper  is a collection of
Hamiltonians $H_m$ and currents $J^{\mu}_m(x)$ that give the same
predictions for bound state, scattering, and electromagnetic
observables.  A preferred interaction can be identified  by requiring
that there are no exchange currents in the charge density.  This view is
certainly supported by historical prejudice and is the basis of most
non-relativistic phenomenology.  This assumption is consistent with
non-relativistic quantum mechanics but it is unfortunately inconsistent
with Poincar\'e invariance, which requires that the charge density satisfy
\cite{Polyzou}:  
\begin{equation} 
[ H,\rho (0)] +i \sum_{l=1}^3 [P^l,[K^l,\rho (0)]] =0 
\label{eq:CA} 
\end{equation}
\begin{equation} 
[ \vec{J}, \rho (0)] = 0 
\label{eq:CB} 
\end{equation} 
\begin{equation} 
[ K^l,[K^l, \rho (0)]] + \rho (0) =0 
\label{eq:CC} 
\end{equation} 
where $H$, $\vec{P}$, $\vec{J}$, $\vec{K}$ are the Poincar\'e
generators. These relations cannot be satisfied by an impulse charge
density in a model with interactions.  This problem cannot be defined
away by a clever unitary transformation. 

In order to understand why a perfectly good Hermitian operator should
not be considered an observable it is useful to consider the difference 
between any two of the representations corresponding to $m'\not=m$. Both
models are fit that same experimental data.  They differ by a unitary
transformation, $W_m^{\dagger}W_{m'}$, which looks like the identity
when the particles are far apart, but has some correlations when they
are close together. If we consider this unitary transformation as a
change of coordinates then  both coordinate systems look the same when
considering the behavior of asymptotically separated free  particles. 
In both models the free Hamiltonian is has the same form  (as it is used
to compute the $S$-matrix).  The $m$ representation  appears to have a
coordinate system with short range correlations  relative to the $m'$
while the $m'$ representation  appears to have a coordinate system with
short range correlations  relative to the $m$ system. The models {\it cannot}
be distinguished based on how well they fit bound state and
scattering observables.   

The problem is that most experiments only measure asymptotic particles. 
There are clearly some limitations to this picture.   Particles are not
measured at infinite distances from the interaction region.   Models
having a deuteron with a macroscopic RMS radius would require a bizarre
reinterpretation of physics; on the other hand without a probe that 
explicitly couples to the strong RMS radius, it is impossible to find an
experimental basis for preferring a model with a specific  RMS radius.
One consequence of this theorem is that it always possible to 
introduce non-localities that will make the RMS radius identical to the 
charge radius. 

The main message in this purely academic exercise is that in the absence
of an interaction that couples directly to the RMS radius,  the RMS 
radius is not a true quantum mechanical observable.  It is a
representation dependent quantity where the ``correct'' choice of
representation cannot be  determined by experiment.  More practical
considerations related to interpreting various radii as observables are
discussed in a recent preprint of Friar, Martoerll, and Sprung
\cite{Friar}.   Derivations of nucleon-nucleon interactions from field
theory models also make specific ``approximations'' which correspond to
different representations in the above sense.  These considerations make
it clear that there is no preferred NN-interaction.   It is fortuitous
that the many-body physics seems to be relatively insensitive to
specific choices of realistic interactions.

\section{Acknowledgements}
The author is grateful to  Fritz Coester, Sid Coon, and Jim Friar 
for comments that contributed materially to this work. 
This work was supported by the Department of Energy, Nuclear
Physics Division, under contract DE-FG02-86ER40286.

\pagebreak
\mediumtext

\end{document}